\documentclass[runningheads]{svmult}

\usepackage{makeidx}   
\usepackage{graphicx}  
\usepackage{subeqnar}  
\usepackage{multicol}  
\usepackage{physprbb}  
\makeindex             

\def\theta{\vartheta}
\def\phi{\varphi}

\def\ltsima{$\; \buildrel < \over \sim \;$}
\def\lsim{\lower.5ex\hbox{\ltsima}}
\def\gtsima{$\; \buildrel > \over \sim \;$}
\def\gsim{\lower.5ex\hbox{\gtsima}}

\begin{document}
\title*{Early afterglows as probes for the reionization epoch}
\titlerunning{Early afterglows as probes for the reionization epoch}
%
\author{Davide Lazzati\inst{1,2} \and Gabriele Ghisellini\inst{1} \and
Francesco Haardt\inst{3} \and Alberto Fern\'andez--Soto\inst{1}}
\authorrunning{D. Lazzati et al.}
%
%
\institute{Osservatorio Astronomico di Brera, Via Bianchi 46
I--23807 Merate (Lc), Italy \and Present Address: University
of Cambridge, Institute of Astronomy, Cambridge CB3 0HA, UK
\and Universit\`a dell'Insubria, via Valleggio 11, I--22100 Como, Italy}

\maketitle              

\begin{abstract}
The nature of Gamma--Ray Burst (GRB) progenitors is still a debated 
issue, but consensus is growing on the association of GRBs with massive stars.
Furthermore, current models for the reionization of the universe
consider massive Pop--III stars as the sources of the ionizing photons.
There could then be a natural link between GRBs and reionization. The
reionization epoch can be measured through prompt IR spectral observations
of high redshift GRBs. 
For this, GRBs are better than quasars: 
they produce a smaller HII region even if they are much brighter 
than quasars (but for a much shorter time) and then, contrary
to quasars, they do not modify the absorption properties of the 
surrounding IGM.
\end{abstract}

\section{Introduction}

The importance of Gamma--Ray bursts (GRB) for cosmological studies
has been realized immediately after the measure of the first
high--redshift bursts (GRB971214: Kulkarni et al., 1998 and GRB 000131:
Andersen et al., 2000). Their importance for cosmological
studies is based on observed properties and on the likely association 
with very massive stars:

\begin{itemize}

\item For a few hours, the early afterglow flux in the NIR, optical and
X--ray bands is larger than the flux from any other cosmological object;

\item The afterglow brightness depends very weakly on redshift,
given its time evolution and spectral properties (Ciardi \& Loeb, 2000);

\item GRBs are probably connected with the death of massive stars 
and can hence be used to investigate the star formation rate at
very high redshift, (see, e.g., Lamb \& Reichart, 2000);

\item GRBs are likely linked to the Pop--III objects
and hence to the reionization of the universe (see, e.g. Rees 2000);

\end{itemize}

We here concentrate on the use of GRBs for the measurement
of the redshift of the reionization of the intergalactic medium (IGM).

This measurement is made difficult by the fact that the opacity 
of the IGM is given by (see, e.g., Madau \& Rees, 2000):
\begin{equation}
\tau(z) = 1.5 \times 10^5\,h^{-1}\,\Omega_M^{-1/2}\,
\left({{\Omega_b\,h^2}\over{0.019}}\right)\,\left({{1+z}\over{8}}
\right)^{3/2}
\end{equation}
implying that at high redshift ($z>2$), even if only a hydrogen atom 
over $10^4$ is neutral, yet the universe has a huge ($\tau \sim 10$) 
opacity to UV photons.

Since all observed quasars have a strong Lyman--$\alpha$ line in emission 
in their spectra, it has been speculated that the discovery of a quasar 
without an emission Lyman--$\alpha$ line proves its location
at a redshift larger than reionization.
On the other hand, Madau \& Rees (2000) have shown that a quasar is always
surrounded by a large HII region. 
This implies that the Lyman--$\alpha$
photons produced by the central source will be redshifted at a lower
frequency before reaching the edge of the HII region, and will not be 
scattered by neutral hydrogen any more. The presence/absence of 
Lyman--$\alpha$ emission line in QSO spectra is then not a probe of the 
reionization epoch. 

We here show a possible solution: a GRB is more luminous than a quasar
but, since its duty cycle is much shorter, the total number of
ionizing photons is smaller and the size of the HII region smaller.

\section{Str\"omgren spheres}

An important advantage of GRBs with respect to QSOs for cosmological
use is the dimension of the Str\"omgren sphere surrounding them.

Consider a source which radiates ultraviolet photons. The recombination 
time of the hydrogen with interstellar medium densities, even at redshift 
of order 10 and in presence of moderate clumping, is longer than 1 Gy 
(Madau \& Rees, 2000).
In these conditions, the radius of the HII region surrounding the 
photon source is obtained by equating the number of hydrogen atoms 
within a certain volume with the number of ionizing photons emitted 
during the life of the source itself. We obtain:
\begin{equation}
R = \left( {{3\,\dot N_{\rm ion}\, t}\over{4\pi\,n_H}} \right)^{1/3},
\label{eq:rstr}
\end{equation}
where $\dot N_{\rm ion}$ is the ionizing photon rate, 
$t$ is the lifetime of the photon source and $n_H$
the hydrogen number density.

By adopting fiducial values for the luminosity and the duty cycle
of a quasar and a GRB, we obtain\footnote{Here and in the following we 
parameterize a quantity $Q$ as: $Q = 10^x\,Q_x$}:
\begin{equation}
R_{\rm QSO} = 1.5\times10^5 \, L_{46}^{1/3}\,t_{15}^{1/3}\, n_H^{-1/3}\,
{\rm pc};
\;\;\;\;\;\;
R_{\rm GRB} = 70 \, L_{46}^{1/3}\,t_{15}^{1/3}\, n_H^{-1/3}
\;\;{\rm pc}
\end{equation} 
There is then a factor 
greater than 1000 between the size of the HII region of a GRB,
(which remains well within a galaxy), 
and the size of the HII region of a QSO. 
In particular, the size of the Str\"omgren sphere
of the quasar is so large that a Lyman--$\alpha$ line emitted
by the central object can travel through an opaque universe and be
observed at infinity. This happens because the line photons
would be redshifted outside the Lyman--$\alpha$ resonance
before reaching the edge of the HII region (Madau \& Rees, 2000; 
Cen \& Haiman, 2000).

\section{Discussion}

We have shown that the early afterglows of GRBs are better suited 
for the study of the properties of the IGM at cosmological distance 
than any other known class of objects. GRBs, in fact, have at the same time
a high luminosity, which makes them easy to detect
and to study, and a small number of emitted photons,
so that their presence do not influence the properties of the
surrounding medium. They are therefore ideal probes, with the smallest
possible impact on what we want to measure. Yet, they are very bright.
It is fair to say that it is still unknown whether or not GRBs emit 
Lyman--$\alpha$ line radiation. What we have shown here is that, 
should such a line be observed in a GRB spectrum, this would imply 
$z_{\rm GRB}<z_{\rm Reion}$, while this is not true for QSOs.

This set of good properties is possible thanks to the 
fact that GRBs are transient
phenomena. For this reason, we must be able to detect and follow up
them in real time in order to fully exploit all the information
they carry.
In particular, it is important to select the high redshift bursts
as soon as possible in order not to waste too much telescope time
on nearby objects. This can be done through prompt NIR imaging,
by the Lyman drop--out technique. For this reason, robotic IR
telescopes are planned to complement the foreseen SWIFT mission
(see, in particular, Zerbi et al. 2001).

\end{document}